\documentclass{ws-procs975x65}

\def\beq{\begin{equation}}
\def\eeq{\end{equation}}

\begin{document}

\title{Embedding Non-Linear Structures in $f(R)$ Cosmologies}
\author{Timothy Clifton}

\address{School of Physics \& Astronomy, Queen Mary University of London,\\
Mile End Road, London E1 4NS, UK\\
E-mail: t.clifton@qmul.ac.uk}

\begin{abstract}
When using Einstein's equations, there exist a number of techniques for embedding non-linear structures in cosmological backgrounds. These include Swiss cheese models, in which spherically symmetric vacua are patched onto Friedmann solutions, and lattice models, in which weak-field regions are joined together directly. In this talk we will consider how these methods work in $f(R)$ theories of gravity. We will show that their existence places constraints on the large-scale expansion of the universe, and that it may not always be possible to consider the Friedmann solutions and weak-field solutions of a theory independently from each other.
\end{abstract}


\bodymatter

\section{Introduction}

There are a number of methods that can be used to consistently include non-linear structures in cosmological models. The most well-known of these is probably the ``Swiss cheese'' approach, in which spherically symmetric solutions are joined to homogeneous and isotropic Robertson-Walker geometries\cite{ES} (see Fig. \ref{fig1}(a)). An alternative approach is to take regions of perturbed Minkowski space, and join them together at reflection symmetric boundaries\cite{L1,L2} (see Fig. \ref{fig1}(b)). The former of these approaches has the benefit of admitting exact solutions. The latter benefits from admitting space-times with reduced symmetry, and from removing the regions of locally homogeneous and isotropic space-time entirely.
\def\figsubcap#1{\par\noindent\centering\footnotesize(#1)}
\begin{figure}[h]
\begin{center}
  \parbox{2.1in}{\includegraphics[width=2.1in]{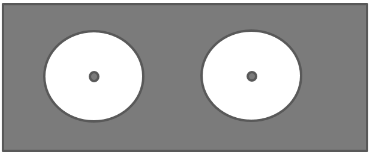}\figsubcap{a}}
  \hspace{4pt}
  \parbox{2.44in}{\includegraphics[width=2.44in]{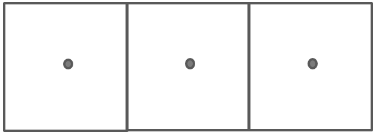}\figsubcap{b}}
  \caption{Two methods of embedding non-linear structures into a cosmological model. (a) A Swiss cheese model. (b) A lattice model. The solid filled region indicates a homogeneous fluid.}
  \label{fig1}
\end{center}
\end{figure}
\vspace{-10pt}

In the context of Einstein's theory, the Swiss cheese and lattice models can be used to elucidate the links between the gravitational fields of non-linear structures, and those of the cosmological models within which they reside. They show that the Friedmann solutions of Einstein's equations are entirely compatible with the Coulomb-like gravitational fields that are expected on small scales. One does not restrict the other. In this paper we will consider the corresponding situation in the $f(R)$ theories of gravity. We will reach quite different conclusions in this case.

\section{The $f(R)$ Theories of Gravity}

The $f(R)$ theories of gravity are a generalisation of Einstein's theory of gravity. They are defined by replacing the Ricci scalar, $R$, in the Einstein-Hilbert action, with some function, $f(R)$. Variation with respect to the metric then gives the following set of field equations:
\beq
\label{fes}
G_{\mu \nu} = \frac{T_{\mu \nu}}{f_R} + \frac{1}{f_R} \left[ \frac{1}{2} (f-R f_R) g_{\mu \nu} + \nabla_{\nu} \nabla_{\mu} f_R - g_{\mu \nu} \nabla_{\alpha} \nabla^{\alpha} f_R \right] \, ,
\eeq
where subscript $R$ denotes a derivative with respect to that quantity, and where $T_{\mu \nu}$ is the energy-momentum tensor of the matter fields (defined in the usual way). These equations are fourth order in derivatives of the metric, and if one want to join together two space-times then the following conditions\cite{jc} are required for their union to be a solution of Eq. (\ref{fes}):
\begin{eqnarray}
\label{jc1}
f_R \left[ \partial_y K^*_{ab} \right]^+_- &=& 0\, , \hspace{40pt} \left[ \partial_y K \right]^+_- = 0 \, , \hspace{40pt} \left[ \partial_y \gamma^*_{ab} \right]^+_- = 0 \, ,\\
f_{RR} \left[ \partial_y R \right]^+_- &=& 0\, , \hspace{40pt} \left[ \partial_y R \right]^+_- = 0\, ,
\label{jc2}
\end{eqnarray}
where $\partial_y$ is the normal derivative to the boundary, $\gamma_{ab}$ is its induced metric, and $K$ and $K^*_{ab}$ are its trace and trace-free parts of the extrinsic curvature, respectively. The brackets $[ ... ]^+_-$ denote the difference in a quantity when evaluated on either side of the boundary, and we have assumed here that there are no surface layers ({\it i.e.} there are no $\delta$-function matter fields on the boundary). The first three of these equations are the same as required by Einstein's theory (up to a factor of $f_R$), while the last two are new. These latter equations constitute additional requirements that a geometry must obey, in order to be a solution of the field equations.

In much of the analysis reported here, we will us a post-Newtonian appoximation to describe weak gravitational fields in the presence on non-linear structures. The leading-order part of the perturbed metric is then given by
\beq
ds^2 \simeq -(1-2 \phi) dt^2 + (1+2 \psi) \delta_{ij} dx^i dx^j \, ,
\eeq
where $\phi$ and $\psi \ll 1$ can be interpreted as gravitational potentials. The Taylor series approximation to theories that can be expanded around $R=0$ is then given by
\beq
\label{anal}
f(R) = f(0) + f_R(0) R + \frac{1}{2} f_{RR}(0) R^2 + O(R^3) \, .
\eeq
If we take $f(0) \sim f_R(0) \sim f_{RR}(0) \sim 1$, then the gravitational potentials become\cite{pn}
\beq
\label{anal2}
\phi = \frac{1}{f_R} \left( \hat{U} + \frac{1}{2} f_{RR} R \right) \hspace{20pt} {\rm and} \hspace{20pt} \psi = \frac{1}{f_R} \left( {U} - \frac{1}{2} f_{RR} R \right) \, ,
\eeq
where $\hat{U}$, $U$ and $R$ are functions that satisfy $\nabla^2 \hat{U} = -4 \pi \rho - f(0)/2$, $\nabla^2 {U} = -4 \pi \rho + f(0)/4$, and $3 f_{RR} \nabla^2 R - f_R R = -8 \pi \rho +2 f(0)$. The reader will note that the first two of these potentials are exactly the same as the potentials that result from solving Einstein's equations with a cosmological constant (where $f(0)$ takes the place of the usual symbol $\Lambda$). The Ricci scalar, $R$, plays the role of an additional Yukawa potential.

\section{Swiss Cheese in $f(R)$ Gravity}

Let us first consider Swiss cheese constructions, in theories that can be expanded as in Eq. (\ref{anal}). If we model the region around a point-like mass using the post-Newtonian approximation, discussed above, then an application of the junction conditions in Eq. (\ref{jc1}) tells us that the radius of the vacuole that surrounds it must obey\cite{paper1}
\begin{eqnarray}
\label{sc1}
\frac{\dot{\Sigma}^2}{\Sigma^2} &\simeq& - \frac{2 U_{,r}\vert_{\Sigma}}{f_R \Sigma} - \frac{k}{\Sigma^2} + \frac{f_{RR}}{f_R} \frac{R_{,r}\vert_{\Sigma}}{\Sigma}\\
\frac{\ddot{\Sigma}}{\Sigma} &\simeq&  \frac{\hat{U}_{,r}\vert_{\Sigma}}{f_R \Sigma} + \frac{f_{RR}}{2 f_R} \frac{R_{,r}\vert_{\Sigma}}{\Sigma} \, ,
\label{sc2}
\end{eqnarray}
where the boundary is at coordinate distance $r=\Sigma$ from the centre of the vacuole, and where $k$ is a constant. The first two terms in each of these equations behave in the same way as the contributions from dust and $\Lambda$ in the Friedmann solutions of Einstein's equations. The second term in Eq. (\ref{sc1}) behaves in the same way as a spatial curvature term, and the last terms in Eqs. (\ref{sc1}) and (\ref{sc2}) have no counterparts in the solutions of Einstein's equations.

In we now impose the remaining junction conditions, from Eq. (\ref{jc2}), then we see that $R_{,r} =0$ on the boundary, such that the last terms in each of Eqs. (\ref{sc1}) and (\ref{sc2}) must vanish. Recall that these are the only two terms that show any difference from the solutions of Einstein's equations, and one is led to the conclusion that the only Swiss cheese models that can be constructed within this class of theories are ones that expand at leading order in a way that is identical to the Friedmann solutions of Einstein's equations with $\Lambda$. This is a strong restriction, imposed on the large-scale expansion simply by demanding that the theory have a well-defined post-Newtonian limit in the vicinity of point-like masses.

\section{Lattice Models in $f(R)$ Gravity}.

The lattice models, while mathematicall and conceptually quite different from the Swiss chees models, display a remarkable similarity in the final result of their construction. In this case it can be shown, as a result of both the field equations and the junction conditions, that the position of the boundary must satisfy\cite{paper1}
\beq
\frac{\ddot{Z}}{Z} \simeq  \frac{\hat{U}_{,z}\vert_Z}{ f_R Z} \, ,
\eeq
where here the boundary is taken to be at coordinate distance $z=Z$ from the centre of the lattice cell. This is again for theories that can be expanded as in Eq. (\ref{anal}). The right-hand side of the equation above takes the same form as the source terms in the Friedmann solutions of Einstein's equation in the presence of dust and $\Lambda$. We therefore have the same result as before: The only large-scale expansion that is possible, in the leading-order behaviour of these constructions, is identical to  that of the Friedmann solutions of Einstein's equations with $\Lambda$.

\section{Non-Analytic $f(R)$ Models}

The results derived so far seem quite limiting, but one might question whether they are a result of assuming the theory can be expanded as in Eq. (\ref{anal}). Another possiblity is to consider non-analytic functions, where this is not possible. The simplest of these would be theories with $f(R)=R^{1+\delta}$. Exact solutions for point-masses surrounded by spherically symetric vacua have been found in Refs.~\refcite{exact1} \& \refcite{exact2}. A static vacuum solution is given by\cite{exact1}
\beq
\label{e1}
ds^2 = -A_1(r) dt^2 +\frac{dr^2}{B_1(r)} +r^2 (d\theta^2 + \sin^2 \theta d \phi^2) \, ,
\eeq
where
\begin{eqnarray*}
A_1(r) &=& r^{\frac{2 \delta (1+2\delta)}{(1-\delta)}} + c_1 r^{-\frac{(1-4 \delta)}{(1-\delta)}} \, ,\\
B_1(r) &=& \frac{(1-\delta)^2}{(1-2 \delta+4 \delta^2) (1-2 \delta -2 \delta^2)} \left(1+c_1 r^{-\frac{(1-2 \delta +4 \delta^2)}{(1-\delta)}} \right) \, .
\end{eqnarray*}
A non-static solution, on the other hand, also exists, and is given by\cite{exact2}
\beq
\label{e2}
ds^2 = -A_2(r) dt^2 + t^{\frac{2\delta(1+2\delta)}{(1-\delta)}} B_2(r) (dr^2 +r^2 (d\theta^2 + \sin^2 \theta d \phi^2)) \, ,
\eeq
where
\begin{equation*}
A_2(r) = \left( \frac{1-\frac{c_2}{r}}{1+\frac{c_2}{r}} \right)^{2/\sigma} 
\hspace{20pt} {\rm and} \hspace{20pt}
B_2(r) = \left( 1+\frac{c_2}{r}\right)^4 A^{\sigma+2 \delta -1} \, .
\end{equation*}
The quantities $c_1$ and $c_2$ that appear in these equations are both arbitrary constants. The only other remaining quantitiy is $\sigma \equiv 1- 2 \delta +4 \delta^2$.

Although the weak-field limit of these non-analytic theories is cumbersome to probe, these exact solutions can be used to try and construct Swiss cheese models. If one attempts this, however, then one finds that no such constructions are possible\cite{paper1}. The junction conditions across the boundary cannot be satisfied unless $\delta=0$, which reduces the theory to that of Einstein. This is true for the geometry described in Eq. (\ref{e1}), and the geometry described in Eq. (\ref{e2}). It is also true of the Schwarzschild solution, which is often another solution of the field equations of $f(R)$ theories. The only exception to this latter case is if $f(R)$ is linear in $R$, which is again just the Einstein equations with $\Lambda$. So, once again, we find that the only large-scale expansion that is possible is the same as that prescribed by the Friedmann solutions of Einstein's equations.

\section{Models of $f(R)$ Dark Energy}

Another family of theories that cannot be expanded as in Eq. (\ref{anal}) are those considered in Refs. \refcite{de1}, \refcite{de2} \& \refcite{de3}. These theories have attracted a lot of attention in the cosmology community, so it seems well worth investigating them within the current context. We start by expanding the quantities $f$ and $f_R$ in terms of their order-of-smallness in the post-Newtonian expansion. This gives
\beq
f= f^{(0)} + f^{(2)} + O(\epsilon^4) \hspace{20pt} {\rm and} \hspace{20pt} f_R =  f_R^{(0)} + f_R^{(2)} + O(\epsilon^4) \, ,
\eeq
where superscripts in brackets denote the order of a quantity in $\epsilon$, which is the post-Newtonian expansion parameter. This is exactly the way that the weak-field limit of these theories is usually treated.

If one now tries to build one of the constructions we considered earlier, then one is immediately led to the following\cite{paper2}:
\beq
\label{first}
f^{(0)} =0 \hspace{20pt} {\rm and} \hspace{20pt} f_R^{(0)} = q(t) \, ,
\eeq
where $q(t)$ is an, as yet unspecified, function of time only. This leads us to consider two possible classes of theory:
\begin{itemlist}
\item {\bf Class I:} These are theories in which $q=$constant. It immediately follows that $f(R) = q R+ F(R)$ in this class, where $F_R \sim \epsilon^2$.
\item {\bf Class II:} These are theories in which $q \neq$ constant. This requires $q=q(R)$, as $f_R^{(0)}=f_R^{(0)}(R)$. For this to be true implies $R$ is a function of time only.
\end{itemlist}
Let us consider each of these classes separately:

\vspace{5pt}
{\bf Class I Theories.} This class of theories contains all of those considered in Refs. \refcite{de1}, \refcite{de2} \& \refcite{de3}. In this case the field equations (\ref{fes}) and junction conditions (\ref{jc1}) give\cite{paper2}
\beq
\label{classi}
\phi = \frac{1}{q} \left( \hat{U} + \frac{1}{2} F_R \right) \hspace{20pt} {\rm and} \hspace{20pt} \psi = \frac{1}{q} \left( U- \frac{1}{2} F_R \right) \, ,
\eeq
where $\nabla^2 \hat{U} = -4 \pi \rho - F(0)/2$ and $\nabla^2 U = -4 \pi \rho + F(0)/4$, and where  $F(0) =$ constant. The reader can once again note that the $\hat{U}$ and $U$ functions are exactly as one would expect from the solutions of Einstein's equations with a cosmological constant (where $F(0)$ now plays the role of what is usually denoted $\Lambda$). The only possible new behaviour comes the the terms containing $F_R$, which must obey\cite{paper2}
\beq
\nabla^2 F_R = \frac{1}{3} R + \frac{2}{3} F(0) - \frac{8 \pi}{3} \rho \, .
\eeq
If, however, we impose the junction conditions from Eq. (\ref{jc2}), then it can be seen that the normal derivative of $F_R$ must vanish on the boundary. This means that the extra terms in Eq. (\ref{classi}) cannot contribute to the motion of the boundary. We therefore find, once again, that the only possible large-scale expansion is indistinguishable from that found in the Friedmann solutions of Einstein's equations with $\Lambda$.

\vspace{5pt}
{\bf Class II Theories.} The leading-order part of $R$, in all theories in this class, must be a function of time only. This immediately implies that the leading-order part of $f^{(2)}$ must also be a function of time only, as $f^{(2)}=f^{(2)}(R)$. If this is true, it is straightforward to show the post-Newtonian parameter $\gamma$ must be given by\cite{paper2}
\beq
\gamma \equiv \frac{\psi}{\phi} = \frac{1}{2} \, .
\eeq
This is strongly inconsistent with observational bounds, and occurs because the fifth-force that occurs in this class of $f(R)$ theories cannot be screened by the Chameleon mechanism. This can be seen from the equation of motion of the effective scalar degree of freedom, which is given by
\beq
\nabla^2 f_R^{(2)} = -\frac{8 \pi}{3} \rho + \, {\rm homogeneous \; terms} \, .
\eeq
There does not exist sufficient freedom to create a non-trivial potential for $f_R^{(2)}$ in this expression, and so the fifth-force goes un-screened. This means that all theories within this class, while potentially admitting interesting new cosmological behaviour, cannot be considered observationally viable.

\section{Conclusions}

We find that, in the presence of non-linear structure, the observationally viable theories we have studied all evolve on large scale like the Friedmann solutions of Einstein's equations with $\Lambda$. We also note that the oscillatory behaviour that is generically present in the Friedmann solutions of many of these $f(R)$ theories\cite{de1} appears to be supressed. This indicates that either the weak-field limit of these theories is destroyed when the oscillations occur, or that the existence of non-linear structures prevents the oscillations from occuring at all. Whatismore, in every viable case, the effective cosmological constant that emerges must be constructed from constant parameters that appear linearly at lowest order in the gravitational Lagrangian ({\it i.e.} just like $\Lambda$ does). This leads us to the conclusion that these theories do not solve any of the problems associated with the cosmological constant, and hence that the difficulties created by introducing a new light degree of freedom on small scales have no obvious benefit.

\vspace{5pt}
\flushleft
{\bf Acknowledgments.} I am grateful to Peter Dunsby, Rituparno Goswami, and Anne Marie Nzioki, who were my collaborators in the scientific work on which this article is based.

\end{document}